\documentstyle[amstex,amsfonts,12pt]{article}

\begin{document}



\thispagestyle{empty} \vspace*{1cm} \rightline{Napoli DSF-T-18/2003} %
\rightline{INFN-NA-18/2003} \vspace*{2cm}

\begin{center}
{\LARGE A twisted conformal field theory description of }

{\LARGE dissipative quantum mechanics }

{\LARGE \ }

\vspace{8mm}

{\large Gerardo Cristofano\footnote{{\large {\footnotesize Dipartimento di
Scienze Fisiche,}{\it \ {\footnotesize Universit\'{a} di Napoli
\textquotedblleft Federico II\textquotedblright\ \newline
and INFN, Sezione di Napoli}-}{\small Via Cintia - Compl.\ universitario M.
Sant'Angelo - 80126 Napoli, Italy}}}, Vincenzo Marotta\footnotemark[1], }

{\large Adele Naddeo\footnote{{\large {\footnotesize Dipartimento di Scienze
Fisiche,}{\it \ {\footnotesize Universit\'{a} di Napoli ``Federico II''
\newline
and INFM, Unit\`{a} di Napoli}-}{\small Via Cintia - Compl. universitario M.
Sant'Angelo - 80126 Napoli, Italy}}} }

{\small \ }

{\bf Abstract\\[0pt]
}
\end{center}

\begin{quotation}
We show how the recently proposed CFT for a bilayer Quantum Hall system at
filling $\nu =\frac{m}{pm+2}$ \cite{cgm2}$-$\cite{noi}, the Twisted Model
(TM), is equivalent to the system of two massless scalar bosons with a
magnetic boundary interaction as introduced in \cite{callan1}, at the so
called \textquotedblleft magic\textquotedblright\ points. We are then able
to describe, within such a framework, the dissipative quantum mechanics of a
particle confined to a plane and subject to an external magnetic field
normal to it. Such an analogy is further developed in terms of the TM
boundary states, by describing the interaction between an impurity with a
Hall system.

\vspace*{0.5cm}



{\footnotesize Work supported in part by the European Communities Human
Potential}

{\footnotesize Program under contract HPRN-CT-2000-00131 Quantum
Spacetime\newpage } \setcounter{page}{2}
\end{quotation}

\section{Introduction}

In ref.\cite{cgm5} the effect of the dissipative term ($\eta \overset{{{{\cdot}%
} }}{q}$) on the motion of an electron confined in a plane and in the
presence of an external magnetic field $B$, normal to the plane, was
analyzed. By using the correspondence principle it was possible to quantize
the system and to study its time evolution on large time scales ($t\gg \frac{%
1}{\eta }$) employing coherent states techniques. It was found that the
effect of dissipation would simply be accounted for by a rotation combined
with a scale transformation on the coordinate $z$ of the electron: $%
z\rightarrow (\rho e^{i\gamma })z$, $\rho =\left( \frac{\eta ^{2}+\omega ^{2}%
}{\omega ^{2}}\right) ^{\frac{1}{2}}$, $\gamma =-\arctan \left( \frac{\eta }{%
\omega }\right) $. As a result, the gaussian width describing the electron
in the lowest Landau level ($LLL$) state would get reduced. Furthermore the
current operator, lying in the Hall direction in the absence of dissipation,
for $\eta \neq 0$ would acquire a longitudinal component, with a resulting
metallic conductance $\sigma _{L}\neq 0$ for the multielectron system.

It is remarkable that such an effect was soon after proposed in ref.\cite%
{callan1} in the context of boundary conformal field theories (BCFT). The
authors consider a system of two massless free scalar fields which have a
boundary interaction with a periodic potential and furthermore are coupled
to each other through a boundary magnetic term. By using a string analogy,
this last interaction allows for exchange of momentum of the open string
moving in an external magnetic field. The magnetic interaction enhances one
chirality with respect to the other producing the same effect of a rotation
together with a scale transformation on the fields as for the dissipative
system of ref. \cite{cgm5} where the string parameter places the role of
dissipation. It is crucial to observe that conformal invariance of the
theory is preserved only at special values of the parameters entering the
action, the so called \textquotedblleft magic\textquotedblright\ points.

The aim of this letter is to show that the effective conformal field theory,
the twisted model (TM), recently proposed in refs. \cite{cgm2}$-$\cite{noi}
is well suited to describe a dissipative system precisely at the
\textquotedblleft magic\textquotedblright\ points. In fact the presence of a
$Z_{m}$ twist accounts, in the open string picture, for a mismatch of
momentum exchange at its two endpoints. For the special $Z_{2}$ case the TM
has central charge $c=2$ and describes a system of two layers coupled
through a topological defect. It is interesting to notice that the fields
which diagonalize such an interaction can be expressed in terms of the
original layers fields through a rotation and a scale transformation. The
amount of the rotation and scale transformation is fixed by the order of
the twist. This is the content of the $m$-reduction procedure introduced in
ref. \cite{cgm1}. The paper is organized as follows:

In sec. 2 we recall some results of the dissipative quantum mechanics
obtained in refs. \cite{cgm5}\cite{callan1}.

In sec. 3 we review the properties of the TM which are relevant to the
description of the dissipative effects and make a correspondence with the
BCFT approach of ref. \cite{callan1}.

In sec. 4 we analyze the effect of the magnetic field on the TM boundary
states (BS) recently introduced in ref. \cite{noi} and make a correspondence
with the bulk degrees of freedom of ref. \cite{cgm4}.

In sec. 5 we give explicitly the duality transformations which relate the UV
properties with the IR ones in the TM context.

\section{Dissipative quantum mechanics}

In the presence of dissipation, the equations of motion of an electron
moving in a plane with a magnetic field $B$ transverse to it are given by:
\begin{equation}
\frac{d^{2}z}{dt^{2}}=-\left( \eta +i\omega \right) \frac{dz}{dt}\text{ \ \
\ \ \ \ \ \ \ \ \ \ \ \ \ \ \ \ \ \ }\frac{d^{2}\overline{z}}{dt^{2}}%
=-\left( \eta -i\omega \right) \frac{d\overline{z}}{dt},
\end{equation}
where $z\left( t\right) =x\left( t\right) +iy\left( t\right) $, $\omega =B$
(in the units $m=c=e=1$) and $\eta $ is the viscosity coefficient.

By assuming the following relations between the canonical momenta $p_{z}$, $%
p_{\overline{z}}$ and the velocities $v_{z}$, $v_{\overline{z}}$ for $\eta
\neq 0$:
\begin{equation}
p_{z}=v_{z}+\frac{\eta +i\omega }{2}z\text{ \ \ \ \ \ \ \ \ \ \ \ \ \ \ \ \
\ \ \ \ }p_{\overline{z}}=v_{\overline{z}}+\frac{\eta -i\omega }{2}\overline{%
z},
\end{equation}%
it is possible to quantize the system using the correspondence principle and
to define the creation and annihilation operators $\widehat{b}$ and $%
\widehat{b}^{+}$
\begin{equation}
\widehat{b}=\frac{\omega +i\eta }{\sqrt{2\omega }}\text{\ }\widehat{%
\overline{z}}_{\infty }^{+}\text{\ \ \ \ \ \ \ \ \ \ \ \ \ \ \ \ \ \ \ \ \ }%
\widehat{b}^{+}=\frac{\omega -i\eta }{\sqrt{2\omega }}\text{\ }\widehat{z}%
_{\infty },
\end{equation}%
with commutation relation $\left[ \widehat{b},\text{\ }\widehat{b}^{+}\right]
=1$. The coordinates $z_{\infty }$, $\overline{z}_{\infty }$ describe the
center of the Larmor orbit of the electron and $\widehat{z}_{\infty }$, $%
\widehat{\overline{z}}_{\infty }$ are the corresponding operators. It is now
interesting to observe that the time evolution on a large time scale ($t\gg
\frac{1}{\eta }$) is simply defined as:
\begin{equation}
\psi _{j,t\rightarrow \infty }^{\eta }\left( z\right) =\lim_{t\rightarrow
\infty }\left\langle z\left\vert U\left( t,0\right) \right\vert \psi
_{j}\right\rangle =\left\langle z_{\infty }|\psi _{j}\right\rangle .
\end{equation}%
We are interested to the case in which $\psi _{j}\left( z\right) $ is the
wave function of an electron in the lowest Landau level $(LLL)$:
\begin{equation}
\psi _{j}\left( z\right) =\sqrt{\frac{\omega }{\pi }}\frac{1}{\sqrt{j!}}%
\left( z\sqrt{\frac{\omega }{2}}\right) ^{j}e^{-\frac{\omega }{4}\left\vert
z\right\vert ^{2}}.  \label{wavef1}
\end{equation}%
In order to build the state $\left\vert z\left( t\right) \right\rangle
_{t\rightarrow \infty }$ it is possible to define the coherent state $%
\left\vert \xi \right\rangle $ such that $\widehat{b}\left\vert \xi
\right\rangle =\xi \left\vert \xi \right\rangle $ where $\xi =\frac{\omega
-i\eta }{\sqrt{2\omega }}\overline{z}$ and $\left\vert \xi \right\rangle $
is given by $\left\vert \xi \right\rangle =e^{-\frac{\left\vert \xi
\right\vert ^{2}}{2}}e^{\widehat{b}^{+}\xi }\left\vert 0\right\rangle $ with
the vacuum state $\left\vert 0\right\rangle =\sqrt{\frac{\omega }{\pi }}e^{-%
\frac{\omega +i\eta }{4}\left\vert z\right\vert ^{2}}$ annihilated by $%
\widehat{b}$: $\widehat{b}\left\vert 0\right\rangle =0$.

If we require that the zero angular momentum state $\psi _{0}\left( z\right)
$ is annihilated by $\widehat{b}$ (that is we require unitarity in our
description), then we immediately get
\begin{equation}
\psi _{j,t\rightarrow \infty }^{\eta }\left( \xi \right) =e^{-\frac{%
\left\vert \xi \right\vert ^{2}}{2}}\left\langle 0\left\vert e^{\widehat{b}%
\overline{\xi }}\frac{\left( \widehat{b}^{+}\right) ^{j}}{\sqrt{j!}}%
\right\vert 0\right\rangle =\sqrt{\frac{\omega ^{2}+\eta ^{2}}{\pi \omega }}%
\frac{1}{\sqrt{j!}}\left( \frac{\omega -i\eta }{\sqrt{2\omega }}z\right)
^{j}e^{-\frac{\omega ^{2}+\eta ^{2}}{4\omega }\left\vert z\right\vert ^{2}},
\label{wavef2}
\end{equation}
having expressed $\xi $, $\overline{\xi }$ in terms of the original $z$, $%
\overline{z}$ variables. By comparing eq. (\ref{wavef2}) with eq. (\ref%
{wavef1}) we can infer that the effect of dissipation can be simply
accounted for by making the following transformations on the variable $z$: $%
z\rightarrow (\rho e^{i\gamma })z$, where $\rho =\left( \frac{\eta
^{2}+\omega ^{2}}{\omega ^{2}}\right) ^{\frac{1}{2}}$ and $\gamma =-\arctan
\left( \frac{\eta }{\omega }\right) $, that is a rotation plus a scale
transformation.

Furthermore for a vector operator $O$ one has the transformation properties:
\begin{eqnarray}
O_{x,t\rightarrow \infty }(z) &=&\frac{1}{\rho }\left( O_{x,0}\cos \gamma
-O_{y,0}\sin \gamma \right)   \nonumber \\
O_{y,t\rightarrow \infty }(z) &=&\frac{1}{\rho }\left( O_{x,0}\sin \gamma
+O_{y,0}\cos \gamma \right) .  \label{vecop1}
\end{eqnarray}%
The striking consequence of the above relations (see ref.\cite{cgm5}) is
that, if one starts with the current density operator $\overrightarrow{J}$
which accounts for a Hall conductance $\sigma ^{H}=\frac{1}{\omega }$ (being
$\left\langle \overrightarrow{J}\right\rangle =\left( -\frac{E}{\omega }%
,0\right) $) only, by applying the transformations given in eq. (\ref{vecop1}
) one obtains
\begin{equation}
\left\langle \overrightarrow{J}^{\eta }\right\rangle =\left( -\frac{\omega }{%
\omega ^{2}+\eta ^{2}}E,\frac{\eta }{\omega ^{2}+\eta ^{2}}E\right)
\label{current}
\end{equation}%
with a resulting metallic conductance $\sigma ^{L}=\frac{\eta }{\omega
^{2}+\eta ^{2}}$ different from zero! The brackets above indicate an
expectation value.

It is remarkable that such an effect was soon after proposed in ref. \cite%
{callan1} in the context of BCFT. A system of two massless scalar fields in $%
1+1$ dimensions is considered, which are free in the bulk except for
boundary interactions, which couple them. Its action is given by $%
S=S_{bulk}+S_{pot}+S_{mag}$ where:
\begin{eqnarray}
S_{bulk} &=&\frac{\alpha }{4\pi }\int_{0}^{T}dt\int_{0}^{l}d\sigma \left(
\left( \partial _{\mu }X\right) ^{2}+\left( \partial _{\mu }Y\right)
^{2}\right) , \\
S_{pot} &=&\frac{V}{\pi }\int_{0}^{T}dt\left( \cos X\left( t,0\right) +\cos
Y\left( t,0\right) \right) ,  \label{SP} \\
S_{mag} &=&i\frac{\beta }{4\pi }\int_{0}^{T}dt\left( X\partial
_{t}Y-Y\partial _{t}X\right) _{\sigma =0}.
\end{eqnarray}%
In the equations above $\alpha $ determines the strength of dissipation and
is related to the potential $V$, as it can be seen, by rescaling the fields;
$\beta $ is related to the strength of the magnetic field $B$ orthogonal to
the $X-Y$ plane, as $\beta =2\pi B$.

The magnetic term introduces a coupling between $X$ and $Y$ at the boundary
keeping conformal invariance. Such a symmetry gets spoiled by the presence
of the interaction potential term except for the magic points $\left( \alpha
,\beta \right) =\left( \frac{1}{n^{2}+1},\frac{n}{n^{2}+1}\right) ,n\in
{\Bbb Z}$. For such parameters values the theory is conformal invariant for
any potential strength $V$. It is possible to express all the degrees of
freedom of such a system in terms of the boundary states, which can be
easily constructed in two steps:

\begin{itemize}
\item By considering first the magnetic interaction term it is easy to see
that the net effect of the magnetic field is a chiral $o(2)$ rotation of the
Neumann boundary state $|N>$ as:
\begin{equation}
|B_{0}>=\sec \left( \frac{\delta }{2}\right) e^{i\delta {\cal R}_{M}}|N>.
\end{equation}

Above the rotation operator ${\cal R}_{M}$ is given by
\begin{equation}
{\cal R}_{M}=(y_{L}^{0}p_{L}^{X}-x_{L}^{0}p_{L}^{Y})+\sum_{n>0}\frac{i}{n}%
\left( \alpha _{n}^{Y}\alpha _{-n}^{X}-\alpha _{-n}^{Y}\alpha _{n}^{X}\right)
\label{rot1}
\end{equation}
and the rotation parameter $\delta $ is defined in terms of the parameters $%
\alpha ,\beta $ as $\tan \left( \frac{\delta }{2}\right) =\frac{\beta }{%
\alpha }$.

\item By considering in addition the effect of the potential term one
obtains the boundary state $|B_{V}>$ as:
\begin{equation}
|B_{V}>=\sec \left( \frac{\delta }{2}\right) e^{i\delta {\cal R}%
_{M}}e^{-H_{pot}\left( 2X_{L}^{^{\prime }}\right) -H_{pot}\left(
2Y_{L}^{^{\prime }}\right) }|N^{X^{^{\prime }}}>|N^{Y^{^{\prime }}}>
\end{equation}
\end{itemize}

where the rotated and rescaled coordinates $X^{^{\prime }},Y^{^{\prime }}$
have been introduced as:
\begin{eqnarray}
X^{^{\prime }} &=&\cos \frac{\delta }{2}\left( \cos \frac{\delta }{2}X-\sin
\frac{\delta }{2}Y\right)  \nonumber \\
Y^{^{\prime }} &=&\cos \frac{\delta }{2}\left( \sin \frac{\delta }{2}X+\cos
\frac{\delta }{2}Y\right) .  \label{XY1}
\end{eqnarray}%
Since the $o(2)$ rotation commutes with the Virasoro generators $L_{n}$, the
state $|B_{V}>$ satisfies the Ishibashi condition $\left( L_{n}-\widetilde{L}%
_{-n}\right) |B_{V}>=0$. Notice the strong resemblance of the above
relations with the transformation properties given in eq. (\ref{vecop1}),
describing the effect of dissipation on the vector operator $O$. We will
come back later to further comment on this analogy, after introducing the TM
in section 3. Let us only observe here that it is possible to get further
insight into this analogy by considering the partition function $Z_{NB_{V}}$
defined as:
\begin{equation}
Z_{NB_{V}}=\sec \left( \frac{\delta }{2}\right) <N|q^{L_{0}+\widetilde{L}%
_{0}}|B_{V}>  \label{part1}
\end{equation}%
because, in the open string language, the rotation ${\cal R}_{M}$ introduces
now twisted boundary conditions in the $\sigma $ direction.

In the following section we will introduce the two interacting layers system
in the picture of the twisted theory and show that the interlayer
interaction is diagonalized by the effective fields $X,\phi $ which are
related to the layers fields $Q^{(1)},Q^{(2)}$ just by the relation given in
eq. (\ref{XY1}) for $\alpha =\beta $. Also a generalized construction of the
partition function $Z_{NB_{V}}$ will be performed in that context.

\section{The TM model}

In order to introduce the TM model let us consider a system of two
interacting parallel layers of $2D$ electrons in a strong perpendicular
magnetic field $B$. The filling factor $\nu ^{(a)}=\frac{1}{2p+2}$ is the
same for the two $a=1$, $2$ layers (balanced system) while the total filling
is $\nu =\frac{1}{p+1}$. The CFT description for such a system can be given
in terms of two compactified bosons $Q^{(a)}$ (with central charge $c=2$)
defined on the single layer \textquotedblleft $a$\textquotedblright .

We review now the construction of the $Q^{(a)}$ fields in the TM description
in its key steps, which will turn out to be also relevant for the analogy we
are proposing. We show how the $m$-reduction procedure is equivalent to the
effect of a magnetic boundary term; in other words the magnetic term can
result in a twist on the neutral field.

\begin{itemize}
\item Starting from the Fubini field
\begin{equation}
Q(z)=q-iplnz+i\sum_{n\neq 0}\frac{a_{n}}{n}z^{-n}
\end{equation}%
compactified on a circle with radius $R^{2}=1/\nu =2(p+1)$ \cite{cgm4}, we
perform the transformation $z\rightarrow e^{i\theta _{j}}z$ and get $%
Q(e^{i\theta _{j}}z)$, where $z=e^{-i\frac{2\pi x}{L}}$ and $\theta _{j}=%
\frac{2\pi j}{m}$, $j=0,...,m-1$.

\item For $m=2$ (which regards the two layers system considered in the
paper) there are two possible values, $\theta _{0}=0,\theta _{1}=\pi $, with
the corresponding fields:
\begin{eqnarray}
Q(z) &=&q-iplnz+i\sum_{n\neq 0}\frac{a_{n}}{n}z^{-n}\equiv Q^{(1)}\left(
z\right)  \label{Qtwisted1} \\
Q(-z) &=&q+\pi p-iplnz+i\sum_{n\neq 0}\frac{a_{n}}{n}\left( -1\right)
^{n}z^{-n}\equiv Q^{(2)}\left( z\right) \text{.}  \label{Qtwisted2}
\end{eqnarray}
\end{itemize}

\begin{itemize}
\item Summing and subtracting, we get the charged field $X(z)$ and the
neutral one $\phi (z)$ which satisfies twisted boundary conditions $\phi
(e^{i\pi }z)=-\phi (z)$. The $X(z)$ and $\phi (z)$ fields, which have the
profound meaning of diagonalizing the interlayer interaction (see ref. \cite%
{cgm4} and following section) can be rewritten in a more enlightening form
as:
\begin{eqnarray}
X(z) &=&\cos (\theta /4)\left( \sin (\theta /4)Q^{(1)}(z)+\cos (\theta
/4)Q^{(2)}(z)\right)  \label{X} \\
\phi (z) &=&\cos (\theta /4)\left( \cos (\theta /4)Q^{(1)}(z)-\sin (\theta
/4)Q^{(2)}(z)\right) .  \label{phi}
\end{eqnarray}%
Such a transformation consists of a scale transformation plus a rotation;
for $\theta =\pi $ the fields $X(z)$ and $\phi (z)$ of the $Z_{2}$ twisted
theory introduced in refs. \cite{cgm2}\cite{cgm4} are obtained and the
transformations above coincide with the transformations given in eqs. (\ref%
{XY1}) for $\delta =\frac{\pi }{2}$.

\item In this context it is possible to express the effect of a boundary
magnetic term as:
\begin{equation}
|B_{0}(\delta )>=\sec (\theta /4)e^{i\delta {\cal R}_{M}}|N(\theta -2\delta
)>  \label{bm1}
\end{equation}

where
\begin{equation}
|N(\theta )>=\sqrt{R}\sum e^{\sum_{n>0}\overline{a}%
_{-n}^{(X)}a_{-n}^{(X)}}e^{\sum_{n>0}\overline{a}_{-n}^{(\phi
)}a_{-n}^{(\phi )}}|w_{X},0>\otimes |w_{\phi },0>
\end{equation}%
and the rotation ${\cal R}_{M},$ defined in eq. (\ref{rot1}), is now given
in terms of $X$ and $\phi $.

The boundary state for the (untwisted) twisted sector in the folded theory
is obtained from the above rotation on the Neumann boundary state when $%
\delta =0,\frac{\pi }{2}$ and can be seen as due to a boundary magnetic term
according to \cite{callan1}.

\item Finally, if we perform the rescaling $z\rightarrow z^{\frac{1}{2}}$, $%
a_{2n+l}\rightarrow \sqrt{2}a_{n+\frac{l}{2}}$, $q\rightarrow \frac{q}{\sqrt{%
2}}$, for $\alpha =\beta $ $\left( \text{that is also for }\eta =\omega
\right) $, we obtain the $X(z)$ and $\phi (z)$ fields in the standard form;
that is the twisted CFT just constructed (see refs. \cite{cgm2}\cite{cgm4})
can be conjectured to represent the correct CFT which describes dissipative
quantum mechanics (DQM) of ref. \cite{cgm5}.
\end{itemize}

The full set of characters and the partition function for this model was
given in \cite{cgm4}. In \cite{noi} the effect of the presence of an
impurity in the system was analyzed by mapping the impurity into a boundary
state by using the well known folding procedure and the boundary partition
function $Z_{AB}$ was given. In the rest of the paper we will try to
convince the reader that such a conjecture is correct even in the context of
boundary CFT. More precisely we will study the effect of the magnetic field
on the BS of the TM model \cite{noi} and construct the correspondent
partition function $Z_{NB_{V}}$ (see eq. (\ref{part1})). In the following
section we will explicitly see how the boundary state so constructed is
related, for the different values of $\delta =0$ or $\delta =\frac{\pi }{2}$%
, to the description of the bulk degrees of freedom of the TM.

\section{Twisted Model and magnetic boundary interaction}

In order to analyze the effect of a magnetic boundary term in the TM we can
resort to the fermionized theory and define a pair of left-moving fermions
as:
\begin{equation}
\begin{array}{ccc}
\psi _{1}=c_{1}e^{\frac{i}{2}\left( Q^{\left( 2\right) }+Q^{\left( 1\right)
}\right) }=c_{1}e^{iX} &  & \psi _{2}=c_{2}e^{-\frac{i}{2}\left( Q^{\left(
2\right) }-Q^{\left( 1\right) }\right) }=c_{2}e^{i\phi }%
\end{array}%
\end{equation}%
where $c_{i}$, $i=1,2$ are cocycles necessary for the anticommutation.

At the first non-trivial \textquotedblleft magic\textquotedblright\ point $%
\alpha =\beta =\frac{1}{2}$ of ref. \cite{callan1} corresponding in our
model to the $m=2$, $p=0$ case it is very simple to obtain the action of the
magnetic boundary term on the Neumann state $|N>$. In fact, separating the
two Dirac fermions into real and imaginary parts, $\varphi _{1}=\psi
_{11}+i\psi _{12}$, $\varphi _{2}=\psi _{21}+i\psi _{22}$, we get four
left-moving Majorana fermions given by $\psi =\left( \psi _{11},\psi
_{12},\psi _{21},\psi _{22}\right) =\left( \cos X,\sin X,\cos \phi ,\sin
\phi \right) $ and a corresponding set of right-moving ones. In this new
language the magnetic term acts only on the fourth Majorana fermion as ${\rm %
R}_{M}=e^{2i\delta }$ where $\delta =0$ ($\delta =\frac{\pi }{2}$) for the
untwisted (twisted) sector of our theory, being its action the identity for
the other components. Now we add a potential term which acts on the Majorana
as:
\begin{equation}
{\rm R}_{P}=\left(
\begin{array}{cccc}
\cos \left( 2V\right)  & -\sin \left( 2V\right)  & 0 & 0 \\
\sin \left( 2V\right)  & \cos \left( 2V\right)  & 0 & 0 \\
0 & 0 & \cos \left( 2V\right)  & -\sin \left( 2V\right)  \\
0 & 0 & \sin \left( 2V\right)  & \cos \left( 2V\right)
\end{array}%
\right) .
\end{equation}%
So the overall rotation of the corresponding fermionic boundary states is $%
{\rm R}={\rm R}_{M}{\rm R}_{P}$.

In terms of ${\rm R}$ the partition function $Z_{AB},$ where $|A>$ is the
Neumann boundary state $|N>$ and $|B>$ is the magnetic-potential BS $|B_{V}>$%
, can be rewritten as:
\begin{equation}
Z_{NB_{V}}\left( \delta ,V\right) =\left\langle N\right\vert e^{-L\left(
L_{0}+\bar{L}_{0}\right) }|B_{V}\left( \delta \right) >=\sqrt{2}\left(
q\right) ^{-2/24}\prod_{n=1}^{\infty }\det \left( {\rm I}+q^{n-\frac{1}{2}}%
{\rm R}\right)
\end{equation}%
where $q=e^{2i\pi \tau }$ and ${\rm I}$ is the identity matrix.

Finally we get:
\begin{equation}
Z_{NB_{V}}(\delta ,V)=\sqrt{2}\frac{\theta _{3}\left( V|\tau \right) }{\eta
\left( \tau \right) }\sqrt{\frac{\theta _{3}\left( V|\tau \right) }{\eta
\left( \tau \right) }}\sqrt{\frac{\theta _{3}\left( \delta +V|\tau \right) }{%
\eta \left( \tau \right) }},  \label{ZUV1}
\end{equation}%
where $\delta =0$ ($\delta =\frac{\pi }{2}$) for the untwisted (twisted)
sector.

On the other hand, choosing $|A>$ to be the vacuum state we can compute the
partition functions $Z_{AB}$ where $|B>$ are all the BS for the TM obtained
in ref. \cite{noi}. In terms of the characters defined in \cite{cgm4} (for $%
i=0,1$ and $f=0,1$)%
\begin{eqnarray}
\tilde{\chi}_{((i,0),f)}(\tau ) &=&\frac{\theta _{3}\left( \tau \right) }{%
\eta \left( \tau \right) }\frac{\theta _{3}\left( \tau \right) +(-)^{f}\sqrt{%
\theta _{3}\left( \tau \right) \theta _{4}\left( \tau \right) }}{2^{3/2}\eta
\left( \tau \right) }+(-)^{i}\frac{\theta _{4}\left( \tau \right) }{\eta
\left( \tau \right) }\frac{\theta _{4}\left( \tau \right) +(-)^{f}\sqrt{%
\theta _{4}\left( \tau \right) \theta _{3}\left( \tau \right) }}{2^{3/2}\eta
\left( \tau \right) };  \nonumber \\
\chi _{(0,0)}^{+}(\tau ) &=&\frac{\theta _{3}\left( \tau \right) }{\sqrt{2}%
\eta \left( \tau \right) }\frac{\sqrt{\theta _{2}\left( \tau \right) \theta
_{3}\left( \tau \right) }}{\eta \left( \tau \right) };\text{ \ \ \ \ \ \ \ \
\ \ \ \ \ \ }\chi _{(1,0)}^{+}(\tau )=\frac{\theta _{2}\left( \tau \right) }{%
\sqrt{2}\eta \left( \tau \right) }\frac{\sqrt{\theta _{3}\left( \tau \right)
\theta _{2}\left( \tau \right) }}{\eta \left( \tau \right) }  \nonumber \\
\tilde{\chi}_{(0)}(\tau ) &=&\frac{\theta _{2}\left( \tau \right) }{\eta
\left( \tau \right) }\frac{\theta _{2}\left( \tau \right) }{2\eta \left(
\tau \right) };  \label{p1}
\end{eqnarray}%
we get, by using eqs.(\ref{ZUV1},\ref{p1}):%
\begin{equation}
\begin{tabular}{l}
$Z_{NB_{V}}\left( \delta =0,V=0\right) =\tilde{\chi}_{((0,0),0)}(-\frac{1}{%
\tau })+\tilde{\chi}_{((1,0),0)}(-\frac{1}{\tau })+\tilde{\chi}_{((0,0),1)}(-%
\frac{1}{\tau })+\tilde{\chi}_{((1,0),1)}(-\frac{1}{\tau });$ \\
$Z_{NB_{V}}\left( \delta =0,V=\frac{\pi }{2}\right) =2\sqrt{2}\tilde{\chi}%
_{(0)}(-\frac{1}{\tau });\ \ \ \ \ \ \ \ \ Z_{NB_{V}}\left( \delta =\frac{%
\pi }{2},V=0\right) =2\chi _{(0,0)}^{+}(-\frac{1}{\tau });$ \\
$Z_{NB_{V}}\left( \delta =\frac{\pi }{2},V=\frac{\pi }{2}\right) =2\chi
_{(1,0)}^{+}(-\frac{1}{\tau })$%
\end{tabular}%
\end{equation}%
We will comment on such results in the following section, resorting to the
duality properties of the TM vacua, which are explicitly evidenced in its
analogy with the one impurity Kondo model.

\section{Duality properties and comments}

Our description adapts very closely to a system of two interacting Luttinger
liquids coupled resonantly through an impurity placed in between. Indeed, as
stated in ref. \cite{Kane1}, the problem of tunneling between two different
quantum Hall states at $\nu =1$ and $\nu =1/3$ can be mapped onto that of
tunneling through a barrier in a $g_{L}=1/2$ Luttinger liquid; in
particular, if the two layers are equivalent the two tunneling barriers are
symmetric. This is the condition for a perfect resonance, where the system
flows to the perfectly transmitting fixed point \cite{Kane}. Equivalently
such a model can be mapped to an anisotropic two channel Kondo problem in
which the occupation of the impurity state corresponds to the state of the
Kondo spin, the two leads are the two channels, the tunneling amplitudes
play the role of the transverse couplings and the scaling dimensions of the
fields are related to anisotropy. So we get:
\begin{equation}
{\cal L}^{UV}\propto V\left( S_{j}^{x}\cos \left( X\left( 0\right) \right)
\cos \left( \phi \left( 0\right) \right) +S_{j}^{y}\sin \left( X\left(
0\right) \right) \cos \left( \phi \left( 0\right) \right) \right)
\label{LUV}
\end{equation}%
where $X$ and $\phi $ coincide with the charged and neutral fields
introduced in eqs. (\ref{X}) and (\ref{phi}) and $S_{j}^{x(y)}$ are $%
su\left( 2\right) _{q}$ impurity spins in the spin $j$ representation. As
pointed out in ref. \cite{fendley} there is a duality between the UV and the
IR regime in the Kondo problem. Indeed let us consider the anisotropic Kondo
model defined by a boundary Lagrangian as in eq. (\ref{LUV}) where $X$, $%
\phi $ are massless bosonic fields. Such an interaction induces a flow from
Neumann (UV) to Dirichlet (IR) boundary conditions as $V$ increases. The
integrability constraint requires the infrared Lagrangian to contain a
single non-trivial term:
\begin{equation}
{\cal L}^{IR}\propto V_{D}\left( S_{j-1/2}^{x}\cos \left( \frac{\tilde{X}%
\left( 0\right) }{g_{L}}\right) \cos \left( \frac{\tilde{\phi}\left(
0\right) }{g_{L}}\right) +S_{j-1/2}^{y}\sin \left( \frac{\tilde{X}\left(
0\right) }{g_{L}}\right) \cos \left( \frac{\tilde{\phi}\left( 0\right) }{%
g_{L}}\right) \right)  \label{LIR}
\end{equation}%
where now $\tilde{X}\left( 0\right) ,\tilde{\phi}\left( 0\right) $ are the
\textquotedblleft dual\textquotedblright\ bosonic fields\footnote{%
The fields $X\left( 0,t\right) ,\phi \left( 0,t\right) ,\widetilde{X}\left(
0,t\right) ,\widetilde{\phi }\left( 0,t\right) $ appearing in eqs. (\ref{LUV}%
), (\ref{LIR}) are expressed in the folded system description as $\left(
left\right) \pm \left( right\right) $ components respectively (see ref. \cite%
{noi} for a more detailed description).} and the spins are in the
representation of spin $j-1/2$. In conclusion the duality results in the
following transformations:
\begin{equation}
\begin{array}{ccc}
g_{L}\rightarrow \frac{1}{g_{L}}, & V\rightarrow V_{D}, &
S_{j}^{x(y)}\rightarrow S_{j-1/2}^{x(y)}.%
\end{array}%
\end{equation}

If we remember the relation $R=2\sqrt{g_{L}}$ \cite{Lin}, we see that
duality maps the compactification radius $R^{2}$ to $4/R^{2}$, a
half-integer spin to an integer one (i.e. no spin) and conversely. At the
same time the weak coupling limit goes to the strong one, Neumann boundary
conditions to Dirichlet ones and a description in terms of Laughlin
quasiholes to one in terms of electrons. It also maps an untwisted sector to
a twisted one ($\delta \rightarrow \delta +\pi /2$) and conversely. In such
a context, the theory without spin is dual to our TM model in which the
presence of a spin-$1/2$ impurity gives rise to twisted boundary conditions
\cite{noi}. In fact the above duality can be explicitly realized as follows.
Let us consider the UV vacuum corresponding to the boundary partition
function $Z_{NB_{V}}(\delta =0,V=\frac{\pi }{2})$; by performing the
transformations: $\tau \rightarrow -\frac{1}{\tau }$ and $\delta
=0\rightarrow \delta =\pi /2$ we find%
\begin{equation}
Z_{NB_{V}}(\delta =0,V=\frac{\pi }{2})\rightarrow Z_{NB_{V}}(\delta _{D}=%
\frac{\pi }{2},V_{D}=\frac{\pi }{2})
\end{equation}%
i.e. the partition function $Z_{NB_{V}}(\delta _{D}=\frac{\pi }{2},V_{D}=%
\frac{\pi }{2})$ corresponding to the IR vacuum.\ The finite renormalized
value of $V_{D}$\ is the consequence of the non-perturbative multi-solitons
corrections (see ref.\cite{callan1}\ for details). Moreover the effect of
the spin-$1/2$ impurity (a quasihole impurity in our case) is to induce
twisted boundary conditions (i.e. $\delta =0\rightarrow \pi /2$). This is
the analog of the non-Fermi liquid behavior of the overscreened Kondo
problem. Moreover, the duality between the two fixed points with twisted
(untwisted) boundary conditions implies that two equivalent descriptions
exist in which the role of the electrons and quasiholes gets exchanged. That
extends the duality between Laughlin quasiparticles and electrons to the
quasiholes which characterize the paired states. In this case the impurity
spin, which is not present in the classic problem of tunneling between edge
states in the fractional Quantum Hall Effect, plays a crucial role, as
discussed before. The impurity spin leads us to define a larger group with
respect to the modular transformations in which it also plays a role: the
\textquotedblleft duality\textquotedblright\ group.

It is useful now to give a physical interpretation of our results in terms
of dissipation. The present framework adapts very well to the description of
an impurity interacting with a thermal bath which is realized in terms of
two kinds of light particles (i.e. the two Quantum Hall Fluids). The
dissipation is given by the kinetic term for the two fluids and a periodic
potential is added. This problem can be solved in the two regimes of strong
and week \textquotedblleft corrugation\textquotedblright\ \cite{saleur} and
flow occurs toward the stable point. When a magnetic term is added for the
impurity, the system drastically changes its properties and a more stable
point with $\alpha =\beta =1/2$ appears. The effect of $\beta \neq 0$ is to
rotate the current $\vec{J}$ getting a \textquotedblleft
metallic\textquotedblright\ component (see eq.(\ref{current})). In the two
layers system this implies that a current flows between the layers due to
the twisted boundary conditions.

Finally we point out that this description can be also applied to a system
of two Branes interacting with strings as it was proposed in \cite{cgm6}.

\end{document}